# Electrowetting of a soap bubble

Steve Arscott

*Institut d'Electronique, de Microélectronique et de Nanotechnologie (IEMN), CNRS UMR8520, The University of Lille, Cité Scientifique, Avenue Poincaré, 59652 Villeneuve d'Ascq, France*

A proof-of-concept demonstration of the electrowetting-on-dielectric of a sessile soap bubble is reported here. The bubbles are generated using a commercial soap bubble mixture - the surfaces are composed of highly doped, commercial silicon wafers covered with nanometre thick films of Teflon®. Voltages less than 40V are sufficient to observe the modification of the bubble shape and the apparent bubble contact angle. Such observations open the way to *inter alia* the possibility of bubble-transport, as opposed to droplet-transport, in fluidic microsystems (e.g. laboratory-on-a-chip) - the potential gains in terms of volume, speed and surface/volume ratio are non-negligible.

steve.arscott@iemn.univ-lille1.fr

It has long been known[1] that electricity can be used to change the shape of a liquid - this effect is called electrowetting.[2] Electrowetting has a variety of modern applications[3-9] ranging from electronic paper[4,5] and energy harvesting[6] to microelectromechanical systems[7,8] and miniaturized chemistry[9] - all these applications focus on the use of liquid *droplets*. In contrast, the use of liquid *films* in such applications would result in reduced volume and time scales along with a considerable increase in the surface/volume ratio – potentially by orders of magnitude – and bringing its welcomed associated advantages. It has long been known that liquid films can be physically deformed by charging[10-18] and recently, non-electrified liquid films have been used for applications in smart materials[19,20] and micro[21-24] and nanotechnologies.[25,26] Here, a proof-of-concept demonstration of the electrowetting of liquid *films* - in the form of millimetre and sub-millimetre sessile soap bubbles – resting on hydrophobic and hydrophilic surfaces is investigated.

Fig. 1(a) shows a sessile soap bubble resting on a solid surface. The apparent contact angle $\theta_b$ of the bubble is seen to depend on the surface wetting and the thickness of the liquid layer $h$ present at the bubble-solid interface.[27] The surface wetting depends of the surfaces energies $\gamma_{lv}$, $\gamma_{sv}$ and $\gamma_{sl}$, i.e. the physical properties of the liquid and the surface, whilst $h$ depends on the initial volume of liquid and liquid drainage from the bubble. In principle, an 'ideal' sessile bubble is formed when $h \ll R$ where $R$ is the radius of curvature of the bubble. Theoretically, as $h/R \to 0$, $\theta_b \to \frac{1}{2}(\cos\theta_l - 1)$, where $\theta_l$ is the contact angle formed between the bubble solution and the surface[27] – experimentally, this has been shown to be true for surfaces ranging from hydrophilic to superhydrophobic.[27] Let us now consider electrowetting of a sessile bubble using an electrowetting-on-dielectric (EWOD) set-up [Fig. 1(b)]. As with a droplet,[2] application of potential $U$ directly to a conducting bubble will result in the increase of the free energy of the system – this energy is stored: (i) mechanically, in terms of the deformation of the bubble and (ii) electrically, in the dielectric layer directly underneath the bubble – assuming a continuous liquid layer is present at the bubble-surface interface.[27] Deformation of the bubble, i.e. changes in the liquid-solid-vapour surface areas and changes in the internal pressure of the bubble, should lead to a modification of the macroscopic contact angle of the bubble from $\theta_b$ to $\theta_{bU}$ as the potential is increased – as is the case for droplet electrowetting.[2] However, one must also consider that the bubble has an extra internal surface which is not present in a droplet EWOD set-up.

In order to form stable bubbles with a lifetime long enough to perform the measurements (10-60s), a solution with three main components is required: pure water, a thickener and an anionic surfactant. The surfactant (e.g. an organosulphate) enables a stable liquid[28] film to form whilst the thickener (e.g. glycerol) increases the viscosity of the mixture; this reduces drainage and prolongs the lifetime of the bubble. A commercially available bubble solution (Pustefix, Germany) was used for the experiments - the main ingredients of this solution are water (91%), thickener (5%), surfactant (1.7%), neutraliser (1%), stabiliser (1.2%) and preserver (<0.1%). The surface tension of the solution was measured to be 28.2 mJ m$^{-2}$ (standard deviation = 0.3 mJ m$^{-2}$) using the pendant drop method.[29,30] A commercial contact angle measurement instrument with its associated software was employed for the measurements (GBX Scientific Instruments, France). The surface tension of the solution is similar to those used in other experiments concerning soap bubbles and films.[10-18] The electrical conductivity of the bubble solution was measured to be 3.77 mS cm$^{-1}$ using a CDM-83 commercial conductivity meter (Radiometer, Denmark) – a KCl (0.1M) solution was used to calibrate the probe. Millimetre and sub-millimetre sized bubbles were generated from the bubble solution for the experiments using a pipette (Bio-Rad, France) having a tip diameter of ~0.5 mm.

Surfaces enabling electrowetting-on-dielectric (EWOD)[2] experiments were fabricated for the study using commercial 3-inch diameter, polished (100) p-type (0.01 Ω cm) silicon wafers (Siltronix, France). The silicon wafers were cleaned and deoxidized using $H_2SO_4/H_2O_2$ and HF based solutions in a controlled cleanroom environment. Ohmic contacts were formed on the rear surface of the silicon wafers using ion implantation and aluminium evaporation. Uniform Teflon® films were formed on the surface of the silicon wafers using spin-coating of TeflonAF 1600 (DuPont, USA) diluted with Fluorinert FC-75 (3M, USA).[31] The thickness of the Teflon® films was measured to be 25.8 (±1.3) nm and 246.5 (±4.4) nm using a surface profile meter (Bruker Corp., USA). The voltage (0-40V) was applied to the bubble using an E3634A DC power supply (Agilent, USA). The voltages were applied by dipping a hypodermic metal needle ($\varphi$ = 300 µm) into the soap bubble. The voltages were ramped slowly at a ramp rate of ~ 1-5 V s$^{-1}$. All surface preparation and experiments were performed in a class ISO 5/7 cleanroom ($T$ = 20°C±0.5°C; $RH$ = 45%±2%). The data was gathered using a commercial Contact Angle Meter (GBX Scientific Instruments, France). The soap bubble solution was measured to have a contact angle of 62.1° (±0.5°) on the Teflon® surfaces.

Fig. 2 shows photographic evidence for electrowetting of sessile soap bubbles using an EWOD set-up. For sessile bubbles having a radius of curvature $R$ ~ 1 mm and a film thickness of ~ 1 µm,[28] the Bond number (Bo = $\rho^* gR^2/2\gamma$) is less than $10^{-3}$ – thus one can assume that the film portion of the bubble to be perfectly spherical. A small Bond number implies that $\theta_b$ can be extracted by accurately measuring the base length and height of the bubble as a function of applied voltage – despite the angle changes being relatively small. Fig. 3 plots the apparent contact angle of the bubbles versus the applied voltage on the different surfaces tested. Values obtained from the experiments are given in the Table. There are several points to note. First, the zero-bias contact angles agree well with the expected contact angles of sessile soap bubbles on hydrophobic and hydrophilic surface.[27] A contact angle of ~107 is observed for a bubble when $h/R$ = 0.146 [Fig. 3(a)]. Second, as with electrowetting of droplets, the thinner the EWOD insulating layers the smaller the required voltage to observe the effect[2] – this is seen by comparing the 25 nm thickness Teflon® results (open circles and triangles) with the 246 nm thickness Teflon® results (open squares and diamonds). Third, for a given insulator thickness, the largest contact angle variations are observed for smaller values of $h/R$ – compare open circles [Fig. 3(a)] with open triangles [Fig. 3(b)] and open squares [Fig. 3(c)] with open diamonds [Fig. 3(d)]. Fourth, the apparent bubble contact angle variations are relatively small (<10°) – even for small values of $h/R$ [Fig. 3(a)]. The insets to Fig. 3 show the cosine of the experimentally measured bubble angle plotted against the applied voltage $U$ squared. Clearly a linear $\cos\theta_b$ vesus $U^2$ relationship – as is seen for droplet electrowetting[2] before the onset of contact angle saturation [Fig. 4] – is not observed.

EWOD experiments were also conducted using droplets of the bubble solution. Fig. 4 shows plots of the electrowetting of the bubble solution on the Teflon® coated surfaces. The value of $\theta_l$ decreases by ~17° and 12° for the bubble solution on the Teflon® films – contact angle saturation[32] begins at 8V and 16V for the thin and thick Teflon® films respectively. The experimental data agrees well with the Young-Lippmann equation[2] using the dielectric thicknesses given above and dielectric constant of 1.92 for the Teflon® films[31] – this is indicated by the dashed lines in Fig. 4.

The behaviour of the charging of soap bubbles and soap films in an electric field has been studied in the past.[10-18] In general, as the external field is increased the film or bubble will deform to have a cone-like appearance,[10] ejecting material[10,11] in the form of smaller charged bubbles or droplets[14] at some critical value of the applied field – increasing the field still further ultimately results in the bubble bursting.[12,14] However, previous studies have not reported an electrowetting effect – as is the case here – in that that the original bubble spread outs on the surface and remains spherical during deformation.

In an effort to understand the experimental results, i.e. the difference between the contact angle change for the bubble [Fig. 3] and the droplet [Fig. 4], we can compare the free energies[2] of a droplet and the bubble. For both the droplet and the bubble, application of a potential $U$ changes the free energy resulting in stored energy which is both mechanical (surface area changes and volume deformation) and electrical (dielectric charging). These stored mechanical and electrical energies can be computed analytically for the droplet and the bubble by considering a simple spherical cap having a constant volume.[2] Fig. 5 shows plots of the stored energy versus contact angle for a droplet and a bubble having dimensions similar to those used for the experiments. The following values were used to calculate the curves: $\gamma_{lv}$ = 28.2 mJ m$^{-2}$, $\gamma_{sl}$ = 1.8 mJ m$^{-2}$, $\gamma_{sv}$ = 15 mJ m$^{-2}$, $\theta_l$ = 62.1°, $\theta_b$ = 107°, bubble film thickness = 1 µm, $V_b$ = 45.3 nL ($r_0$ = 2 mm), $V_d$ = 8.5 µL, $\varepsilon_r$ = 1.92, $d$ = 250 nm. The curves are obtained by calculating the changes in surface areas $\Delta A_{lv}$, $\Delta A_{sl}$ and $\Delta A_{sv}$, the stored electrical energy $\varepsilon_r\varepsilon_0 U^2/2d$ (per m$^2$) [Fig. 1(b)] and the energy due to a change of the internal pressure, $\Delta(\Delta pV)$. The following assumptions are made: (i) the droplet and the bubble shapes are perfect spherical caps, (ii) the thickness of the bubble film is small ($h<<R$) but finite, (iii) there are no losses (temperature changes and electrical losses – i.e. due to dielectric breakdown - are not considered) and (iv) contact angle saturation[32] is not considered. Fig. 5 clearly demonstrates that to deform a droplet to a given contact angle $\theta$ starting from an initial contact angle $\theta_0$ requires less energy than for a bubble. Despite the fact that $V_b<<V_d$, the extra internal surface of the bubble means that the contact angle variation (for a given voltage) is less for a bubble than a droplet. The inset to Fig. 5 shows the ratio of $E_b/E_d$. If we consider the bubble, for a contact angle close to 107° $E_b/E_d = 2.8$ - i.e. more energy is required to deform the bubble than the droplet - this difference in energy is apparent in the experimental results. As $h/R$ is increased, the system can no longer considered to be a sessile bubble on a solid surface but rather a sessile bubble on a liquid. In this case the droplet spreads out but little change in the bubble contact angle would be expected – this is seen in the experiments.

The author thanks Gérard Cambien (Ecole Centrale de Lille) for help with the solution conductivity measurement and Frank Hein (Pustefix) for discussions concerning the bubble solution.

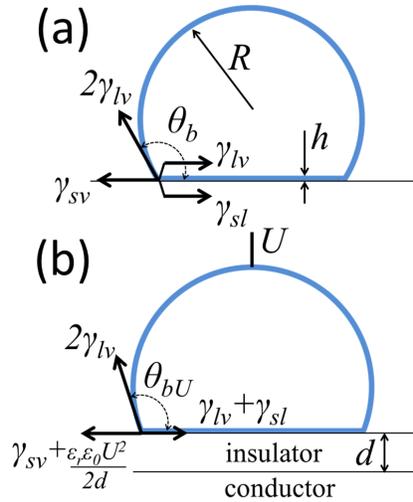

FIG. 1. A sessile bubble in contact with a solid surface (a) at equilibrium and (b) if a voltage is applied to the bubble (considered to be conducting).

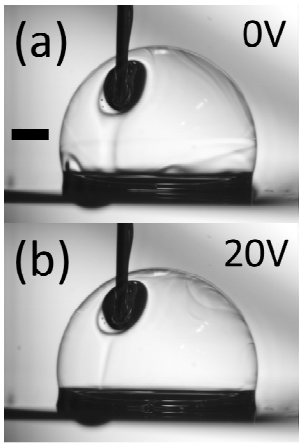

FIG. 2. Experimental evidence of electrowetting of a sessile soap bubble. A bubble resting on a Teflon® (25 nm) covered silicon wafer at (a) 0V and (b) at 20 V. Scale bar = 1000 µm.

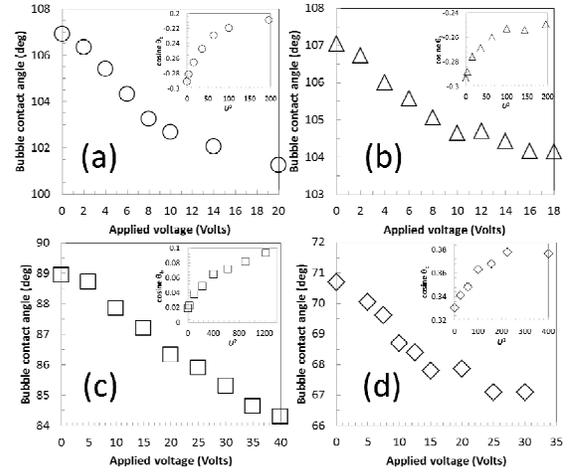

FIG. 3. Plots of the apparent bubble contact angle versus applied voltage. (a) For a Teflon® (25 nm) covered silicon wafer with $h/R$ = 0.146, (b) for a Teflon® (25 nm) covered silicon wafer with $h/R$ = 0.185, (c) for a Teflon® (245 nm) covered silicon wafer with $h/R$ = 0.47 and (d) for a Teflon® (245 nm) covered silicon wafer with $h/R$ = 0.74. The insets show pots of $\cos\theta_b$ versus $U^2$.

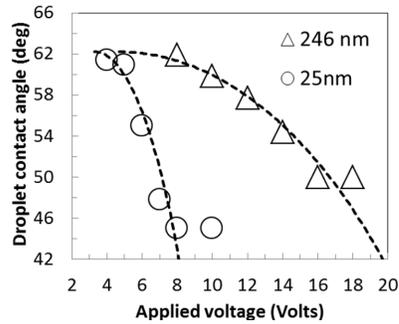

FIG. 4. Electrowetting of a droplet of the bubble solution on the Teflon® coated silicon surfaces (25 nm and 246 nm). The dashed lines are solutions of the Young-Lippmann equation.[2]

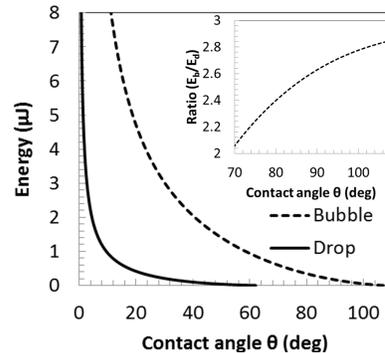

FIG. 5. (a) The free energy of a bubble and a droplet as a function of contact angle. (b) The ratio of the free energies as a function of contact angle.

| Surface | U (Volts) | θb (deg) | b (μm) | ΔP (Pa) | h/R |
|---|---|---|---|---|---|
| Teflon (20 nm) | 0 | 106.9 | 4758 | 45.4 | 0.146 |
| | 20 | 101.2 | 4950 | 44.7 | 0.144 |
| Teflon (20 nm) | 0 | 107.0 | 4680 | 46.1 | 0.185 |
| | 18 | 104.2 | 4824 | 45.3 | 0.182 |
| Teflon (245 nm) | 0 | 88.9 | 4035 | 55.9 | 0.470 |
| | 40 | 84.3 | 4170 | 53.8 | 0.399 |
| Teflon (245 nm) | 0 | 75.6 | 3990 | 54.8 | 0.647 |
| | 40 | 74.0 | 4047 | 53.6 | 0.597 |
| Teflon (245 nm) | 0 | 70.7 | 4023 | 52.9 | 0.739 |
| | 40 | 67.1 | 4158 | 50.0 | 0.623 |

TABLE. Results of the bubble electrowetting experiments.